\documentclass[11pt,a4paper]{article}
\usepackage[T1]{fontenc}
\usepackage[utf8]{inputenc}
\usepackage{geometry}
\usepackage{amssymb}
\usepackage{natbib}
\usepackage{amsmath}
\usepackage{stackrel}
\usepackage{dsfont}
\usepackage{bm}
\usepackage{hyperref}
\usepackage{graphicx}
\graphicspath{ {./images/} }
\usepackage{subcaption}
\usepackage{longtable}
\usepackage{float}
\usepackage{listings}
\usepackage[ruled,vlined]{algorithm2e}
\usepackage{enumitem}
\usepackage{float}

\newcommand{\bear}{\begin{eqnarray}}
\newcommand{\eear}{\end{eqnarray}}
\newcommand{\be}{\begin{equation}}
\newcommand{\ee}{\end{equation}}
\newcommand{\bi}{\begin{itemize}}
	\newcommand{\ei}{\end{itemize}}

\providecommand{\keywords}[1]
{
	\small	
	\textbf{\textit{Keywords---}} #1
}

\title{R package SamplingStrata: new developments and extension to Spatial Sampling}

\author{Marco Ballin\footnote{Italian National Institute of Statistics. Email: ballin@istat.it}, Giulio Barcaroli\footnote{Independent consultant. Email: gbarcaroli@gmail.com}}
\date{}

\begin{document}
	
	\maketitle
	
	\begin{abstract}
		\noindent 
		The R package SamplingStrata was developed in 2011 as an instrument to optimize the design of stratified samples. The optimization is performed by considering the stratification variables available in the sampling frame, and the precision constraints on target estimates of the survey \citep{BB13}. The genetic algorithm at the basis of the optimization step explores the universe of the possible alternative stratifications determining for each of them the best allocation, that is the one of minumum total size that allows to satisfy the precision constraints: the final optimal solution is the one that ensures the global minimum sample size. One fundamental requirement to make this approach feasible is the possibility to estimate the variability of target variables in generated strata; in general, as target variable values are not available in the frame, but only proxy ones, anticipated variance is calculated by modelling the relations between target and proxy variables. In case of spatial sampling, it is important to consider not only the total model variance, but also the co-variance derived by the spatial auto-correlation. The last release of SamplingStrata enables to consider both components of variance, thus allowing to harness spatial auto-correlation in order to obtain more efficient samples.
		
	\end{abstract}
	
	
	\section{Introduction}
	\label{intro}
	
	The R package \textit{SamplingStrata} \citep{barcaroli:2020} is an instrument for the optimization of stratified sampling. It has been described in detail in \cite{GB14}, \cite{BB16} and \cite{barcaroli:2020a} \footnote{Detailed information on the package is in  \url{https://barcaroli.github.io/SamplingStrata/} }. 
	The first release of the R package \textit{SamplingStrata} was made available on the CRAN in July 2011. At the time, the package was being used only internally in the Italian Institute of Statistics (Istat), in particular for agricultural surveys. Since then, as far as we know, \textit{SamplingStrata} has been used in the New Zealand Statistical Institute, tested at Statistics Denmark and considered for evaluation at Statistics Canada. Eurostat used \textit{SamplingStrata} for designing its 2018 LUCAS survey \citep{ballin:2018}. Bank of Italy, supported by Istat, adopted \textit{SamplingStrata} to re-design its Survey on Households Income and Wealth. Also World Bank adopted \textit{SamplingStrata} and embedded it in its SurveySolutions SamplingTools integrated application. Very recently, the NOAA Alaska Fisheries Science Center adopted SamplingStrata for its survey on fish populations in the Gulf of Alaska.
	
\begin{figure}[h!] 
		\centering
		\includegraphics[width=0.6\textwidth]{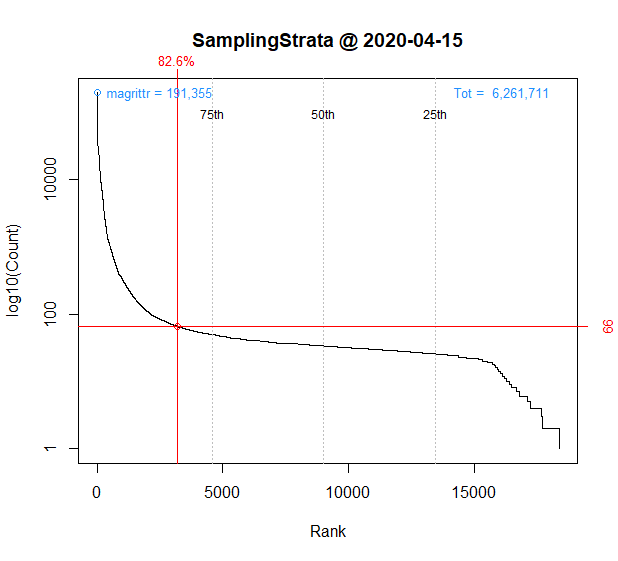}
		\caption{Rank of SamplingStrata in terms of downloads from the CRAN}
		\label{SamplingStrataRank}
	\end{figure}

	Figure \ref{SamplingStrataRank} illustrates the relative importance among all packages in the CRAN: rank in terms of downloads is such that SamplingStrata comes before the 82\% of the about 18,000 packages, despite its highly specific character.
	
	This interest is also due to the improvements that in last two years have been produced by new developments aimed to 
	\begin{itemize}
		\item increase the processing performance of the optimization step;
		\item extend the applicability of the instrument.
	\end{itemize}
	
	With regard to performance, in many occasions the processing time required to perform the optimization step in \textit{SamplingStrata} has been a point of weakness of the package. Compared to the seconds taken by package \textit{stratification} to perform the same task, the minutes (or even hours) required by \textit{SamplingStrata} have been a relevant problem for users. 
	The answers to this problem were:
	\begin{enumerate}
		\item the adoption of a variant of the Genetic Algorithm at the basis of function \textit{optimizeStrata}, i.e. the Grouping Genetic Algorithm;
		\item the development of a completely new optimization algorithm, which does not operate by aggregating atomic strata, but by randomly cutting the domain of each (continuous) stratification variable.
	\end{enumerate}
	
	As for the applicability, the following extensions have been implemented in the package:
	
	\begin{enumerate}
		\item the possibility to model the relationships between auxiliary variables available in the frame and the variables of interest, in order to correctly estimate the variance of the latter (\textit{anticipated variance});
		\item the opportunity to consider the geographical localization of the units in the frame and hence to take into account their \textit{spatial auto-correlation}, in order to correctly determine the best stratification in \textit{spatial sampling}.
	\end{enumerate}
	
	The increasing availability of georeferenced or geocoded sampling frames makes this second point extremely important. 
	For instance, in the Italian Statistical Institute this availability is guaranteed by the \textit{Registro Statistico di Base dei Luoghi, RSBL} (Base Statistical Register of Places). We remind that RSBL offers different levels of geolocation of the statistical units: punctual (coordinates), irregular polygons (sections, municipalities, etc.), regular polygons (EU grids) accompanied by the corresponding geographic files necessary for determining spatial relations.
	
	In order to extend the applicability of \textit{SamplingStrata} to spatial sampling using the methodology suggested by \cite{degruijter2015}, two additional functionalities have been implemented consisting in:
	
	\begin{enumerate}
		\item to extend the possibility to handle the anticipated variance to cases in which a \textit{Spatial Linear Model} is applicable (\textit{full parametric approach});
		\item to make available a different approach that requires as input for each point both prediction and prediction errors (\textit{non parametric approach}).
	\end{enumerate}
	
	In the \textit{parametric} approach, the parameters of a Spatial Linear Model are explicitly supplied, while in the \textit{non parametric} approach only the parameter related to the \textit{range} of spatial correlation must be supplied.
	In both cases, geographical coordinates have to be supplied for each unit in the frame.
	
	In order to ensure continuity with previous applications, the old optimization functions (as \textit{optimizeStrata} and \textit{optimizeStrata2} remain in the package, but a new generic optimization function, namely \textit{optimStrata}, has been made available, with a specific parameter, \textit{method}, that indicates the type of algorithm that will be used in the optimization step:
	\begin{enumerate}
		\item method = "atomic" implies the use of the old algorithm implemented in \textit{optimizeStrata};
		\item method = "continuous" implies the use of the algorithm applicable in case of all continuous stratification variables;
		\item method = "spatial" implies the use of the new algorithm implementing the non parametric approach for spatial sampling.
	\end{enumerate}
	
	This paper is structured as follows.
	In Section \ref{optimization} a description of improvements in the optimization step is given, with some details on the adoption of the Genetic Grouping Algorithm to increase the performance of the already available optimization step, the development of the new function \textit{optimizeStrata2} covering the optimization step, and the handling of anticipated variance.
	In Section \ref{spatial} the general approach followed to extend the applicability of the package to spatial sampling is illustrated.
	In Section \ref{case1} is reported an application of the non parametric approach to a widely used example (dataset \textit{meuse}).
	Section \ref{case2} reports an application of both approaches to the case of optimization of the stratification of a frame where the selection units are Census Enumeration Areas.
	Finally, in Section \ref{conclusions} some conclusions are given, together with indications on future developments.
	
	The R scripts together with data necessary to replicate the two case studies are available at the link \url{https://github.com/barcaroli/R-scripts}.
	
	The version of \textit{SamplingStrata} including developments here illustrated is downloadable from the link
	\url{https://github.com/barcaroli/SamplingStrata}.

	\section{Improvements in the optimization algorithms}
	\label{optimization}

	The initial version of the \textit{SamplingStrata} optimization function (\textit{optimizeStrata}) (now \textit{optimStrata} with method = "atomic") was based on the use of a plain Genetic Algorithm (GA). The starting point was the construction of the \textit{atomic stratification}, obtained as the Cartesian product of the stratification variables, all rigorously categorical (if continuous, a discretization step was required). Every possible given stratification was obtained as an aggregation of atomic strata. In this setting, each stratification is represented by a vector of labels, whose length is equal to the number of atomic strata: atomic strata that share the same label in the vector are collapsed in an aggregate stratum. For each stratum (atomic or aggregate) mean and standard deviation of target variables are calculated by considering their values (or the values of related proxy variables) in the sampling frame. For each stratification, required sample size (and related cost) is calculated by applying the Bethel algorithm \citep{bethel:1989}, under given precision constraints on target estimates. In the GA approach, each vector representing a stratification is considered an (\textit{individual}) in a (\textit{generation}), and its (\textit{fitness}) is measured by the associated sample cost. The usual GA operators (\textit{mutation}) and (\textit{crossover}) are applied in order to produce a new generation from the current one. After a given number of iterations, the best solution represents the stratification with the associated minimum sample cost, compliant with precision constraints.
	
	\subsection{Changes to GA operators \textit{selection} and \textit{crossover}}
	
	The optimization step previously described is formalized in Algorithm \ref{algorithm1}.
	\\
	
	\begin{algorithm}[H]
		\SetAlgoLined
		\SetKwData{Left}{left}
		\SetKwData{This}{this}\SetKwData{Up}{up}
		\SetKwFunction{Union}{Union}
		\SetKwFunction{FindCompress}{FindCompress}
		\SetKwInOut{Input}{input}
		\SetKwInOut{Output}{output}
		\Input{sampling frame, precision constraints, atomic strata}
		\Output{optimal stratification of the sampling frame}
		\BlankLine
		\nlset{REM} generate first population pop[1,]\;\label{generate first}
		\For{$j\leftarrow 1$ \KwTo $popSize$} {
			generate randomly a vector pop[1,j] of integers (from 1 to maxstrata)\;
			aggregate atomic strata using pop[1,j] vector\;
			calculate sample.cost(pop[1,j]) applying Bethel algorithm on resulting aggregate strata\;
			store pop[1,j] with min(sample.cost) as bestSolution\;
		}
		\nlset{REM} generate next populations pop[i,]\;\label{generate next}
		\For{$i\leftarrow 2$ \KwTo $iterations$} {
			apply selection and crossover operators to population pop[i-1,]\;
			generate individuals in pop[i,]\;
			apply mutation operator to pop[i,]\;
			\For{$j\leftarrow 1$ \KwTo $popSize$} {
				aggregate atomic strata using pop[i,j] vector\;
				calculate sample.cost(pop[i,j]) applying Bethel algorithm on resulting aggregate strata\;
				\lIf{sample.cost(pop[i,j] $<$ sample.cost(bestSolution)}
				{bestSolution $\leftarrow$ pop[i,j]}
			}
		}
		\caption{Function 'optimStrata' algorithm (method = "atomic")}
		\label{algorithm1}
	\end{algorithm}
	
	As previously mentioned, using the conventional GA operators the convergence towards the best solution can be exceedingly slow. 
	
	Following some of the indications contained in \cite{oluing:2017} and \cite{oluing:2019}, two of the three GA operators were modified, i.e. \textit{selection} and \textit{crossover}, accordingly to the Grouping Genetic Algorithm (GGA) approach (the \textit{mutation} operator remained unchanged). 
	As for \textit{selection}, instead of selecting parents for breeding completely at random, a probability of selection has been defined by means of a fitness function, in order to privilege most promising individuals.
	As for \textit{crossover}, instead of mixing chromosomes in an undifferentiated way, groups of them in one parent (representing already aggregated strata) are attributed to the other parent when generating a child, preserving their composition. 
	
	The introduction of these innovations since release 1.2 allowed to substantially increase the performance of the optimization function.
	
	\subsection{New optimization algorithm}
	
	We recall that the version of optimization so far described is applicable only to categorical stratification variables: were they not categorical, they have to be discretized, obtaining categorical ordinal variables.
	A criticism of survey managers is that strata in the optimal solution are of difficult interpretation,  as values of stratification variables in a given stratum in general are not contiguous. This problem is relevant when stratification variables are of the continuous type: as already said, they have to be discretized, and there is no warranty that in each resulting optimized stratum the values of this variables are contiguous, i.e. without "holes".
	\\
	
	In order to overcome this problem, a new optimization algorithm, still making use of the Genetic Algorithm, has been implemented.
	This new version, first implemented in the function \textit{optimizeStrata2}, now in \textit{optimStrata} with method = "continuous", is applicable only when all the stratification variables are of the continuous type. Instead of building the atomic strata previously to the optimization step and proceeding to aggregate them, with this new function an individual in a population (a given stratification) is generated by randomly cutting the domains of definition of each stratification variable, and by producing the stratification by applying the Cartesian product to the so obtained discretized values. Then, the optimization proceeds as formalized in Algorithm \ref{algorithm2}.
	Under this new approach, the genoma of a solution is given by the sequence of cuts for the different stratification variables. 
	\\
	
\begin{figure}[H] 
		\centering
		\includegraphics[width=1.0\textwidth]{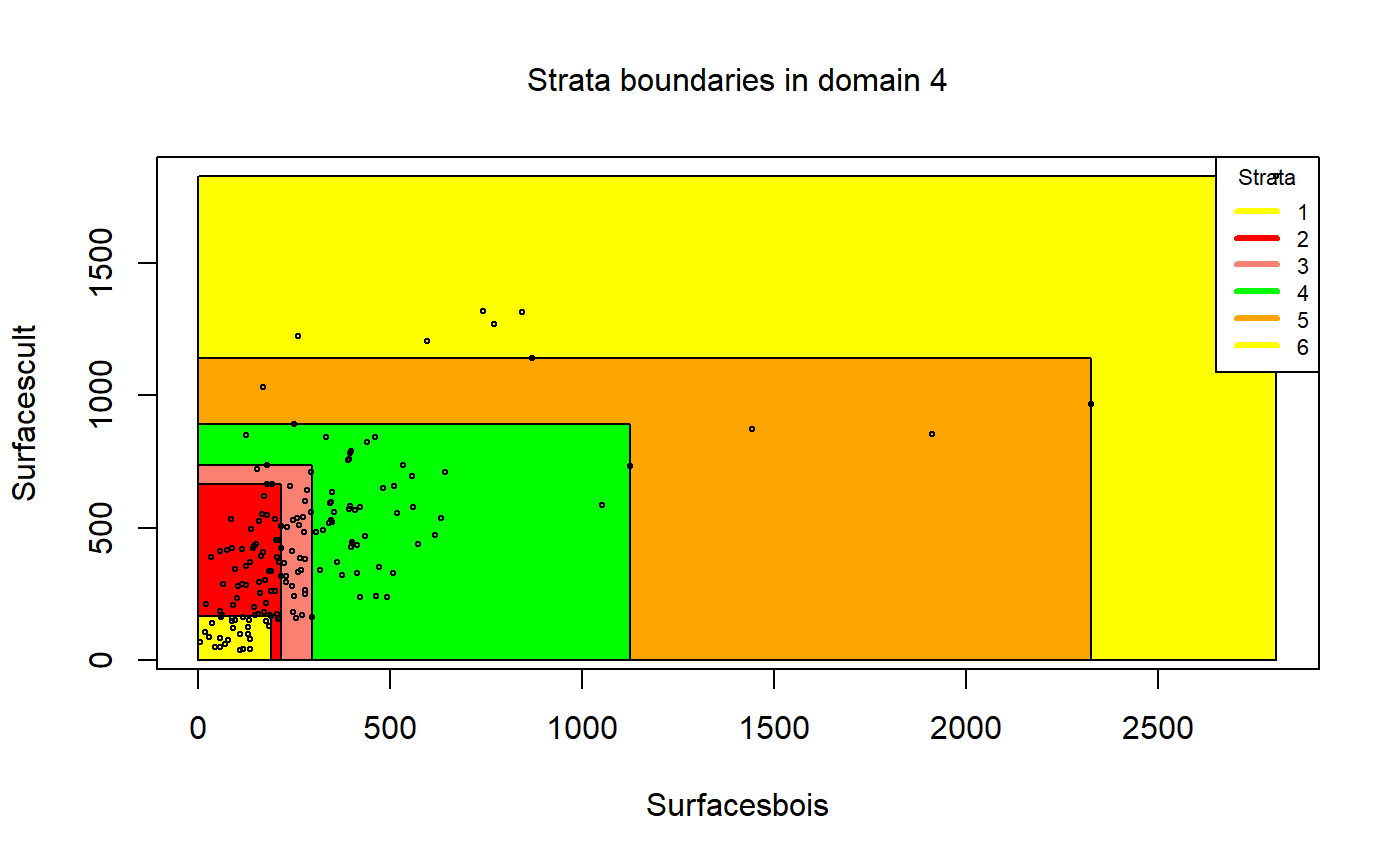}
		\caption{Example of 7-shaped strata obtained with \textit{continuous} method}
		\label{7_shape}
	\end{figure}

	As a result, the obtained strata are "7-shaped" like the ones reproduced in Figure  \ref{7_shape}.
	
	It can be seen that strata are characterized by given non overlapping intervals of values for each stratification variable. 
	\\
	Moreover, in many situations the convergence to the optimal solution can be much quicker with this new algorithm than with the previous one. The suggestion is to try both and analyze the different results in order to choose the best.
	
	\begin{algorithm}[H]
		\SetAlgoLined
		\SetKwData{Left}{left}
		\SetKwData{This}{this}\SetKwData{Up}{up}
		\SetKwFunction{Union}{Union}
		\SetKwFunction{FindCompress}{FindCompress}
		\SetKwInOut{Input}{input}
		\SetKwInOut{Output}{output}
		\Input{sampling frame, precision constraints}
		\Output{optimal stratification of the sampling frame}
		\BlankLine
		\nlset{REM} generate randomly first population pop[1,]\;\label{generate first}
		\For{$j\leftarrow 1$ \KwTo $popSize$} {
			\For{$k\leftarrow 1$ \KwTo $numVar$} {
				generate randomly n cuts (n = nStrata -1) in the definition domain of stratification variable k\;
				concatenate the n cuts to the vector pop[1,j]\;
			}
			generate strata using pop[1,j] vector\;
			calculate sample.cost(pop[1,j]) applying Bethel algorithm on resulting aggregate strata\;
			store pop[1,j] with min(sample.cost) as bestSolution\;
		}
		\nlset{REM} generate next populations pop[i,]\;\label{generate next}
		\For{$i\leftarrow 2$ \KwTo $iterations$} {
			apply selection and crossover operators to population pop[i-1,]\;
			generate individuals in pop[i,]\;
			apply mutation operator to pop[i,]\;
			\For{$j\leftarrow 1$ \KwTo $popSize$} {
				generate strata using pop[i,j] vector\;
				calculate sample.cost(pop[i,j]) applying Bethel algorithm on resulting aggregate strata\;
				\lIf{sample.cost(pop[i,j] $<$ sample.cost(bestSolution)}
				{bestSolution $\leftarrow$ pop[i,j]}
			}
		}
		\caption{Function 'optimStrata' algorithm (method = "continuous")}
		\label{algorithm2}
	\end{algorithm}
	
	\subsection{Anticipated variance}
	
	When optimizing the stratification of a sampling frame, it is assumed that the values of the target variables Y’s are available for the generality of the units in the frame, or at least for a sample of them, by means of which it is possible to estimate means and standard deviations of Y’s in atomic strata. Of course, this assumption is seldom expected to hold. A much more common situation is the one where Y's are proxy variables of the real target variables (named Z). In these cases, there is no guarantee that the final stratification and allocation can ensure the compliance to the set of precision constraints for the real target variables. The importance of considering anticipated variance has been clearly evidenced in \cite{lisic2018}.
	
	In order to take into account this problem, and to limit the risk of overestimating the expected precision levels of the optimized solution, it is possible to carry out the optimization by considering, instead of the expected coefficients of variation related to proxy variables, the anticipated coefficients of variation (ACV) that depend on the model that is possible to fit on couples of real target variables and proxy ones. 
	
	In the current implementation, only models linking continuous variables can be considered. The definition and the use of these models is the same that has been implemented in the package stratification \citep{baillargeon:2014}. In particular, the reference here is to two different models, the linear model with heteroscedasticity:
	\\
	
	$ Z = \beta Y + \epsilon $
	\\
	
	where
	\\
	
	$ \epsilon \sim\ N(0, \sigma^2 Y^{2 \gamma}) $
	\\
	
	(in case $\gamma$ = 0 the model is homoscedastic)
	\\
	
	and the loglinear model:
	\\
	
	$ Z = exp(\beta log(Y) + \epsilon)  $
	\\
	
	where
	\\
	
	$ \epsilon \sim\ N(0, \sigma^2) $
	\\
	
	In the case of the linear model the anticipated variance (AV) on a total is given by the following expression \citep{lisic2018}:
	\\
	
	$AV(\hat{T}_{j}) = \sum_{h=1}^{H} (1- \dfrac{n_{h}}{N_{h}})  \dfrac{N_{h}^2}{n_{h}} ( \sum_{i=1}^{N_{h}} Y_{i,j}^{2\gamma} \sigma^{2} + \sum_{i=1}^{N_{h}} (y_{i}\beta_{j} - \Bar{y}_{h}\beta_{j})^{2} ) $ 
	\\
	
	Using a linear model, in case of presence of heteroscedasticity ($\gamma > 0$) in residuals distribution, a crucial point is in correctly quantifying it. The problem of how to estimate a measure of heteroscedasticity has been dealt with by different authors (\cite{henry:2006} and \cite{knaub:2019}). A simple and quick solution has been implemented in a new function \textit{computeGamma} in \textit{SamplingStrata}. This function receives in input the distribution vectors respectively of residuals and of the explanatory variable, and yields a heteroscedasticity index value together with the value of model variance to be used as values of corresponding parameters. An example of application is in \cite{barcaroli:2020b}.
	
	\section{Extension to spatial sampling}
	\label{spatial}
	
	Let us suppose we want to design a sample survey with $k$ $Z$ target variables, each one of them correlated to one or more of the available $Y$ frame variables. 
	
	When frame units are georeferenced of geocoded, the presence of spatial auto-correlation can be investigated. This can be done by executing for instance the Moran test on the target variables: if we accept the null hypothesis in the Moran test (absence of spatial auto-correlation) then we can search for the optimal stratification with the method based on the calculation of anticipated variance, described in the previous section, or other methods offered by the package functions.
	
	If the null hypothesis is rejected then we have to proceed in a different way taking into account also this variance component.
	
	Two alternative instruments to handle spatial auto-correlation have been implemented in \textit{SamplingStrata}:
	
	\begin{enumerate}
		\item a multivariate version of the methodology proposed by \cite{degruijter2015} has been implemented in a completely new optimization function, i.e. \textit{optimizeStrataSpatial};
		\item in handling anticipated variance, a new model type has been introduced, namely the \textit{Spatial Linear Model}: parameters of this model can be given as input to the \textit{optimizeStrata2} function, that provides to calculate variance in the strata taking into account also the spatial component.
	\end{enumerate}
	
	The first method (the one we call \textit{non parametric}) requires that predicted values together with prediction error variance must be provided for each unit in the frame, by using any kind of model. 
	
	The second (called \textit{parametric}) requires the explicit indication of the parameters of a \textit{Spatial Linear Model} linking target Z and Y frame variables:
	\\
	
	$$Z = Y \beta_{1} + W Y\beta_{2} + u Y^{\gamma} $$
	\\
	
	where the residuals $u$ can be auto-correlated, and the autocovariance is assumed to be exponential.
	$W$ is the \textit{weighted contiguity or distance matrix}. The values obtained as product of the Y values and W matrix must be given for each unit in the frame, together wit the values of the parameters of the model.
	\\
	Moreover, this method explicitly handles the heteroscedasticity $\gamma$ in residuals.	
	\\

	Both methods consider the spatial component in the estimation of the variance in the strata by using the formulas illustrated in \cite{degruijter2015} and \cite{degruijter2019}.
	
	In case $Z$ is the target variable, omitting as negligible the \textit{fpc} factor, the sampling variance of its estimated mean is:
	
	\begin{equation} \label{eq1}
	V(\hat{\bar{Z}}) = \sum_{h=1}^{H}(N_{h}/N)^{2} S_{h}^{2}/n_{h}
	\end{equation}
	
	We can write the variance in each stratum $h$ as:
	
	\begin{equation} \label{eq2}
	S_{h}^{2} = \dfrac{1}{N_{h}^{2}} \sum_{i=1}^{N_{h-1}}\sum_{j=i+1}^{N_{h}}(z_{i}-z_{j})^{2}
	\end{equation} 
	
	The optimal determination of strata is obtained by minimizing the quantity $O$:
	
	\begin{equation} \label{eq3}
	O = \sum_{h=1}^{H} \dfrac{1}{N_{h}^{2}} \{ \sum_{i=1}^{N_{h-1}}  \sum_{j=i+1}^{N_{h}} (z_{i}-z_{j})^{2}\}^{1/2}
	\end{equation}
	
	Obviously, values $z$ are not known, but only their predictions, obtained by means of a regression model. So, in Equation \ref{eq3} we can substitute $(z_{i}-z_{j})^{2}$ with 
	
	\begin{equation} \label{eq5}
	D_{ij}^{2} = \dfrac{(\tilde{z}_{i}-\tilde{z}_{j})^{2}}{R^{2}} + V(e_{i}) + V(e_{j}) - 2Cov(e_{i},e_{j})
	\end{equation}
	
	where $R^{2}$ is the squared correlation coefficient indicating the fitting of the regression model, and $V(e_{i})$, $V(e_{j})$ are the model variances of the residuals. The spatial auto-correlation component is contained in the term $Cov(e_{i},e_{j})$.
	\\
	
	The joint optimization of both strata bounds and sample units allocation can be carried out by software \textit{Ospats} (available at the link \url{https://github.com/jjdegruijter/ospats}) 
	minimizing the expression defined in Equation \ref{eq3} by means of an iterative re-allocation obtained by repeatedly transferring grid points from one stratum to another once verified that these transfers have a positive effect in terms of reduction of the overall variance.
	\\
	
	In particular, the quantity $D_{ij}$ is calculated in this way:
	
	\begin{equation} \label{eq6}
	D_{ij}^{2} = \dfrac{(\tilde{z}_{i}-\tilde{z}_{j})^{2}}{R^{2}} + (s_{i}^{2} + s_{j}^{2}) - 2 s_{i}  s_{j}  e^{-k (d_{ij}/range)}
	\end{equation}
	
	where $ d_{ij} $ is the Euclidean distance between two units i and j in the frame (calculated using their geographical coordinates), and $s_{i}^{2}$ is an estimate of the prediction variance in the single point and \textit{range} is the maximum distance below which spatial auto-correlation can be observed among points. The value of \textit{range} can be determined by an analysis of the spatial \textit{variogram}.
	\\
	
	To summarize, when frame units can be geo-referenced, the proposed procedure is the following:
	
	\begin{enumerate}
		\item acquire coordinates of the geographic location of the units in the population of interest;
		\item fit a model between the Y variable(s) available in the frame and the Z variable(s) of interest;
		\item verify the presence of spatial correlation by executing the Moran auto-correlation test on residuals;
		\item if the test ascertains the presence of spatial correlation, perform a spatial analysis based on the variogram in order to determine the values of the parameters of interest for next steps (in particular, the value of \textit{range});
		\item apply one or both of the two available methods to perform optimization and determine both best stratification and allocation;
		\item when the \textit{non parametric} method is used (with the function \textit{optimizeStrataSpatial}), preliminary steps are:
		\begin{enumerate}[label=\alph*)]
			\item fit a \textit{Kriging} model (or any other spatial model) on data for each couple $Y$ and $Z$ and obtain predicted values together with associated errors for each unit in the frame;
			\item associate predicted values and prediction errors to each unit in the frame;  
		\end{enumerate}	
		\item instead, when the \textit{parametric} method is used (with the function \textit{optimizeStrata2}), preliminary steps are:
		\begin{enumerate}[label=\alph*)]
			\item build the spatial \textit{weighted contiguity matrix} $W$ for each $Y$;
			\item fit a \textit{Spatial Linear Model} for each couple $Y$ and $Z$;
			\item indicate its parameters in the model information frame;
			\item associate values $Y \times W$ for each $Y$ to units in the frame;
			\item in case of heteroscedasticity (verified with a Breusch-Pagan test) calculate the value of related index.
		\end{enumerate}
	\end{enumerate}

	\section{Applications}
	\label{applications}
	
	In the following, two case studies will be illustrated.
	\\
	
	In the first one it will be shown that \textit{optimizeStrataSpatial} and \textit{Ospats} are substantially equivalent in the univariate case, while \textit{optimizeStrataSpatial} can be used also in the multivariate (and multidomain) more general case. 
	\\
	
	In the second case study three different optimization strategies will be compared, differentiating by making use of: 
	\begin{enumerate}
		\item the anticipated variance calculated by means of a simple linear model without spatial auto-correlation (\textit{optimizeStrata2} with model type = 'linear');
		\item use of kriging model with the \textit{non parametric} method (\textit{optimizeStrataSpatial});
		\item the anticipated variance calculated by means of a Spatial Linear Model (\textit{parametric} method with \textit{optimizeStrata2}).
	\end{enumerate}
	
	In the case studies we will make use of a number of R packages specifically dealing with spatial modeling, namely \textit{gstat}, \textit{sp} and \textit{automap}. For a general description of this packages see \citep{pebesma2004}, \citep{pebesma2005}, \citep{graler2016} and \citep{himiestra2008}. 
	
	\subsection{Case study 1: Meuse river territory}
	\label{case1}
	
\begin{figure}[H] 
		\centering
		\includegraphics[width=0.8\textwidth]{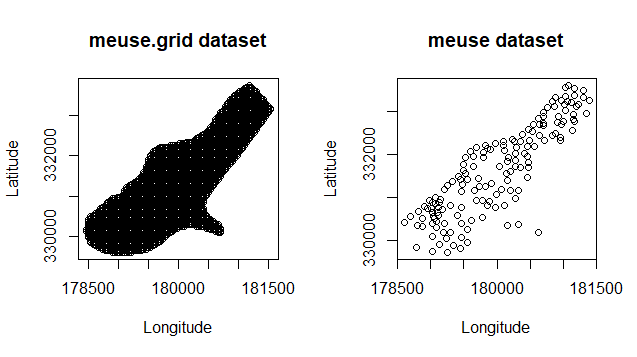}
		\caption{Distribution of meuse.grid and meuse points}
		\label{meuse}
	\end{figure}

	The two datasets "meuse.grid" and "meuse" are provided by the R package \textit{gstat}. 
	We can consider "meuse.grid" as the frame of 3,103 points (each one of approximately 15 m x 15 m) in a flood plain along the river Meuse in the Netherlands, from which a simple random sample of 155 points have been selected in order to observe the concentration of the four metals (see Figure \ref{meuse}).
	
	Dataset "meuse" contains 155 measurements of four heavy metals (cadmium, copper, lead, zinc) measured in the top soil, together with some covariates (elevation, distance from the river, soil type, land use) and coordinates x and y.
	
	Dataset "data.grid" contains basic information (coordinates x and y plus distance from the river and soil type) on a grid of 3,103 points laying in the same zone.
	
	The objective is to produce estimates on the total amount of metals concentrations in the whole area, by adopting an optimized stratified sampling design.
	In order to do that, the following steps are to be performed:
	
	\begin{enumerate}
		\item making use of the available random sample of already observed points, estimate a model for each target variable (metal concentration);
		\item predict values of metal concentration for each point in the area;
		\item determine the best stratification of the frame in order to minimize the size of the sample required to be compliant to precision constraints on the target estimates.
	\end{enumerate}
	
	In the following, we will first consider the univariate case, where only one target estimate will be taken into consideration. This will enable us to compare two different approaches, the one implemented in \textit{SamplingStrata}, the other one implemented in the software \textit{Ospats}.
	
	After showing the equivalence of the two methods, we will consider the multivariate case, where only \textit{SamplingStrata} is applicable.
	
	We consider first the univariate case, and select "lead concentration" as the only target variable.	
	
\begin{figure}[H] 
		\centering
		\includegraphics[width=0.4\textwidth]{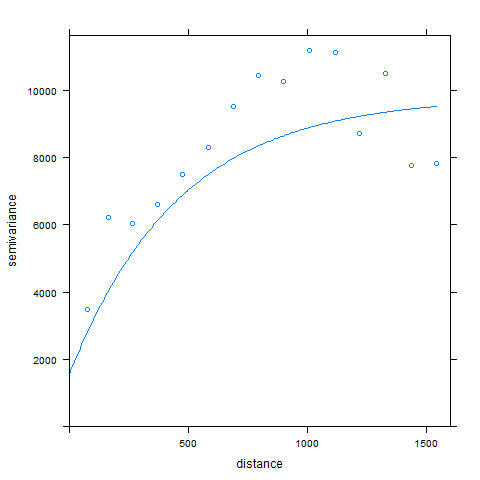}
		\caption{Variogram of lead concentration}
		\label{variogram_lead}
	\end{figure}

	First, we use \textit{kriging} in order to model the relations bewteen \textit{lead concentration} with the covariates \textit{dist} and \textit{soil}. It is a three-step process as suggested by \cite{pebesma:2019}:
	\begin{enumerate}
		\item first we initialize the parameter \textit{psill}, \textit{range} and \textit{nugget} with the function \textit{autofitVariogram} (see the obtained variogram in Figure \ref{variogram_lead}), 
		\item then we fit the model by using the \textit{gstat.lmc} and \textit{fit.lmc} functions;
		\item finally, we predict the the values of \textit{lead concentration} in all the 3,103 points.
	\end{enumerate}
	
	\begin{lstlisting}[language=R]
	library(automap)
	library(gstat)
	data(meuse)
	coordinates(meuse) = c("x", "y")
	data(meuse.grid)
	coordinates(meuse.grid) = c("x", "y")
	v <- variogram(lead ~ dist + soil, data=meuse)
	fit.vgm <- autofitVariogram(lead ~ elev + soil, meuse, 
	model = c("Exp"))
	plot(v, fit.vgm$var_model)
	fit.vgm$var_model
	# model    psill    range
	# 1   Nug 1524.895   0.0000
	# 2   Exp 8275.431 458.3303
	g <- gstat(NULL,"Lead", lead ~ dist + soil, meuse)
	g
	vm <- variogram(g)
	vm.fit <- fit.lmc(vm, g, vgm(psill=fit.vgm$var_model$psill[2], 
	model="Exp", range=fit.vgm$var_model$range[2], 
	nugget=fit.vgm$var_model$psill[1]))
	preds <- predict(vm.fit, meuse.grid)
	names(preds)
	# [1] [1] "Lead.pred" "Lead.var"
	\end{lstlisting}
	
	We are now able to perform the optimization step with \textit{SamplingStrata}, having set a precision constraint equal to 5\% on the Cv of the mean of \textit{lead concentration}:
	
	\begin{lstlisting}[language=R]
	df <- NULL
	df$Lead.pred <- preds@data$Lead.pred
	df$Lead.var <- preds@data$Lead.var
	df$dom1 <- 1
	df <- as.data.frame(df)
	df$id <- meuse.grid$id
	df$lon <- meuse.grid$x
	df$lat <- meuse.grid$y
	frame <- buildFrameSpatial(df=df, 
	id="id", 
	X=c("Lead.pred"), 
	Y=c("Lead.pred"), 
	variance=c("Lead.var"), 
	lon="lon", 
	lat="lat", 
	domainvalue="dom1")
	
	set.seed(1234)
	solution <- optimStrata(method="spatial",
	errors=cv, 
	framesamp=frame,
	nStrata = 3,
	fitting = 1,
	range = fit.vgm$var_model$range[2],
	kappa = 1,
	gamma = 0)
	\end{lstlisting}
	
	\begin{figure}[H] 
		\centering
		\includegraphics[width=0.5\textwidth]{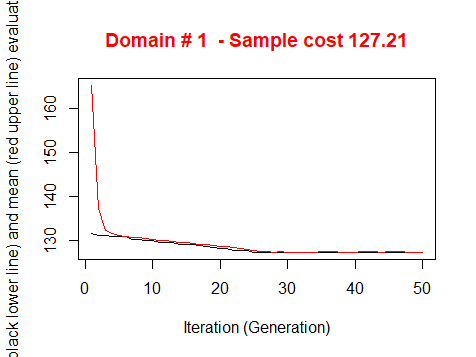}
		\caption{Optimization step for lead concentration with 3 strata}
		\label{lead_optim}
	\end{figure}
	
	In listing~\ref{strata} the structure of the optimized strata is reported. The total sample size is 128.
	
\begin{lstlisting}[float=h,frame=tb,caption=\textit{SamplingStrata} optimized strata,label=strata]
  Domain Stratum Population Allocation SamplingRate Lower_X1 Upper_X1
1      1       1       1360         53     0.038906   0.0000 101.6446
2      1       2       1073         43     0.039848 101.7265 198.1577
3      1       3        670         32     0.047223 198.4284 502.9274
\end{lstlisting}

\begin{lstlisting}[float=h,frame=tb,caption=\textit{Ospats} optimized strata,label=ospatsStrata]
  Domain Stratum Population Allocation SamplingRate Lower_X1 Upper_X1
1      1       1       1370         54     0.039416   0.0000 103.9439
3      1       3       1065         43     0.040376 101.9914 198.7357
2      1       2        668         32     0.047904 197.5290 502.9274
\end{lstlisting}
	
	If we run Ospats on the same input, we obtain a solution that, suitably elaborated in order to be comparable, is reported in listing~\ref{ospatsStrata}. It can be seen that strata are slightly overlapping (when the strata obtained by SamplingStrata are never). The total sample size is 129, practically equivalent to the one determined by \textit{SamplingStrata}. Also the allocations in the strata are almost the same.
	\\
	
\begin{figure}[h!]
		\begin{subfigure}{.5\textwidth}
			\centering
			\includegraphics[width=1.2\linewidth]{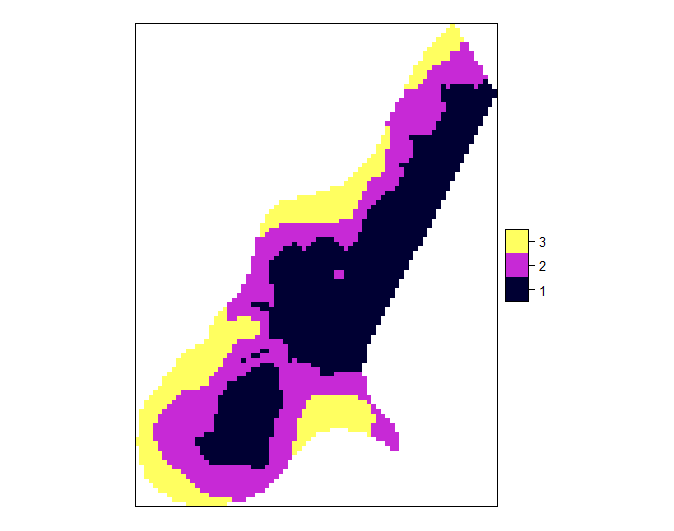}
			\caption{\textit{SamplingStrata} stratification (3 strata)}
			\label{solutions:sfig1}
		\end{subfigure}%
		\begin{subfigure}{.5\textwidth}
			\centering
			\includegraphics[width=1.2\linewidth]{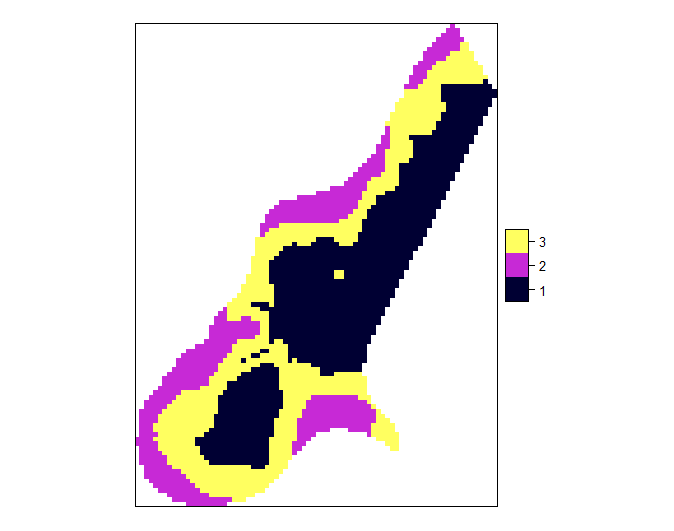}
			\caption{\textit{Ospats} stratification (3 strata)}
			\label{solutions:sfig2}
		\end{subfigure}
		\begin{subfigure}{.5\textwidth}
			\centering
			\includegraphics[width=1.2\linewidth]{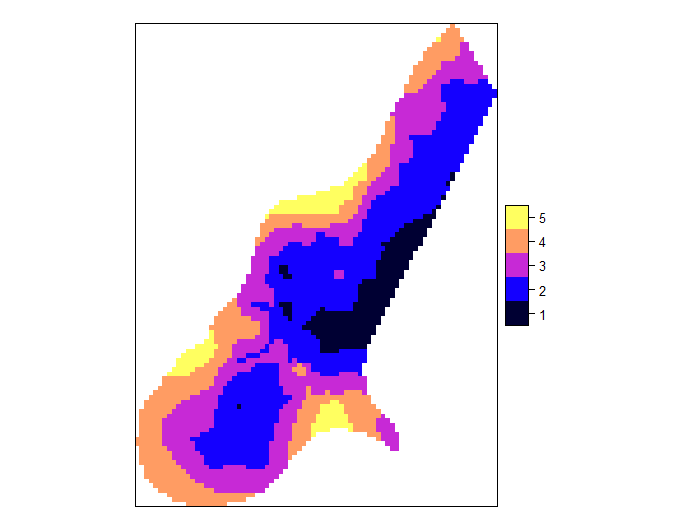}
			\caption{\textit{SamplingStrata} stratification (5 strata)}
			\label{solutions:sfig3}
		\end{subfigure}%
		\begin{subfigure}{.5\textwidth}
			\centering
			\includegraphics[width=1.2\linewidth]{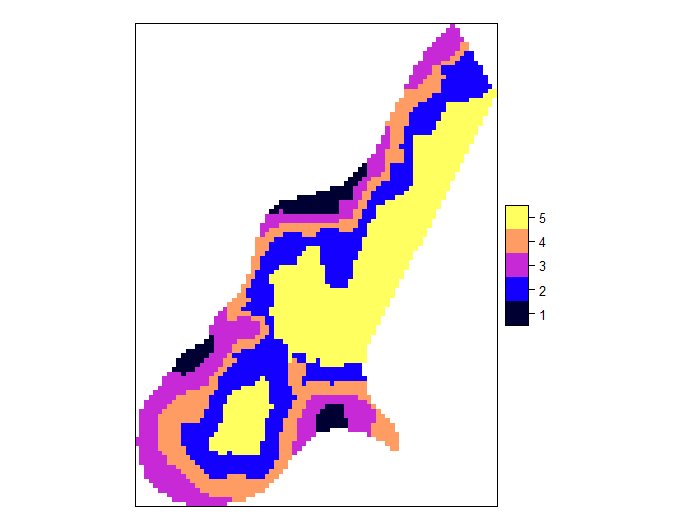}
			\caption{\textit{Ospats} stratification (5 strata)}
			\label{solutions:sfig4}
		\end{subfigure}
		\caption{Comparison of \textit{SamplingStrata} and \textit{Ospats} stratifications}
		\label{solutions}
	\end{figure}

	We run both softwares a second time with a different number of strata (5). Increasing the number of strata does not change the substantial equivalence of the two approaches in terms of allocation (it is 112 for SamplingStrata and 115 for Ospats), but the resulting stratifications are noticeably different, as it can be seen by Figure \ref{solutions}.
	\\

	The extension from the univariate to the multivariate case is straightforward. We now consider four different target estimates, i.e. the total concentration of all metals (cadmium, copper, lead and zinc) in the Meuse area. 
	After producing variograms and models for each one of the target cariables, we are able to predict their concentration for each point in the grid, together with the error prediction, obtaining this work dataframe:
	
	\begin{lstlisting}[language=R]
	df <- NULL
	df$cadmium.pred <- preds.cadmium$cadmium.pred
	df$cadmium.var <- preds.cadmium$cadmium.var
	df$copper.pred <- preds.copper$copper.pred
	df$copper.var <- preds.copper$copper.var
	df$lead.pred <- preds.lead$lead.pred
	df$lead.var <- preds.lead$lead.var
	df$zinc.pred <- preds.zinc$zinc.pred
	df$zinc.var <- preds.zinc$zinc.var
	# df$dom <- meuse.grid@data$soil
	df$dom <- 1
	df <- as.data.frame(df)
	df$id <- meuse.grid$id
	head(df)
	# cadmium.pred cadmium.var copper.pred copper.var lead.pred lead.var 
	# 7.917987     7.007012    65.64426    331.6897   265.9674  7043.944  
	# 8.308433     6.482721    69.01170    308.4855   267.2160  6403.047  
	# 7.898212     6.661317    66.93037    319.6262   260.5370  6574.794  
	# 7.281627     6.828875    63.61440    326.8032   248.9653  6770.737  
	# 8.824362     5.748581    74.34104    245.6440   269.2219  5744.181  
	# 8.333077     6.009188    71.27280    278.8897   261.5889  5909.198  
	# zinc.pred zinc.var 
	# 898.3354  67350.56
	# 918.6911  59349.98
	# 877.8272  61860.84
	# 818.0511  64442.07
	# 949.7147  49761.03
	# 899.6006  52859.14
	\end{lstlisting}
	
	We can now proceed with the optimization step in \textit{SamplingStrata}. As preliminary steps we prepare the precision constraints and sampling frame dataframes:
	
	\begin{lstlisting}[language=R]
	cv
	# DOM  CV1  CV2  CV3  CV4 domainvalue
	# 1 DOM1 0.05 0.05 0.05 0.05           1
	frame <- buildFrameSpatial(df=df, 
	id="id", 
	X=c("cadmium.pred","copper.pred","lead.pred","zinc.pred"),
	Y=c("cadmium.pred","copper.pred","lead.pred","zinc.pred"),
	variance=c("cadmium.var","copper.var","lead.var","zinc.var"), 
	lon="lon", 
	lat="lat", 
	domainvalue="dom")
	\end{lstlisting}
	
	The parameter \textit{range} is now a vector of values:
	
	\begin{lstlisting}[language=R]
	range <- c(fit.vgm.cadmium$var_model$range[2],
	fit.vgm.copper$var_model$range[2],
	fit.vgm.lead$var_model$range[2],
	fit.vgm.zinc$var_model$range[2])
	\end{lstlisting}
	
	In order to "explore" what would be a suitable number of strata in the final solution, we execute the function \textit{KMeansSolutionSpatial}, that applies a kmeans-based clustering method in order to find a quick solution in correspondence of different numbers of strata, starting from 2 ending to a 10:
	
	\begin{lstlisting}[language=R]
	kmeans <- KmeansSolutionSpatial(frame,
	fitting=c(1,1,1,1),
	range=range,
	kappa=1,
	gamma=0,
	errors=cv,
	maxclusters = 10)
	\end{lstlisting}
	
	\begin{figure}[H] 
		\centering
		\includegraphics[width=0.7\textwidth]{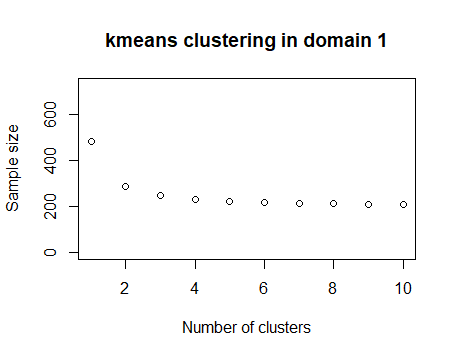}
		\caption{Kmeans solutions up to 10 strata}
		\label{meusekmeans}
	\end{figure}
	
	As shown in Figure \ref{meusekmeans} there is no sensible reduction in sample size after 5 strata, so we choose to set nStrata equal to 5 in the optimization step.	
	
	\begin{lstlisting}[language=R]	
	set.seed(4321)
	solution <- optimStrata (
	method="spatial",
	errors=cv, 
	framesamp=frame,
	iter = 50,
	pops = 10,
	nStrata = 5,
	fitting = c(1,1,1,1),
	range = range,
	kappa = 1,
	writeFiles = FALSE,
	showPlot = TRUE,
	parallel = FALSE
	)
	\end{lstlisting}
	
	\begin{figure}[H] 
		\centering
		\includegraphics[width=0.5\textwidth]{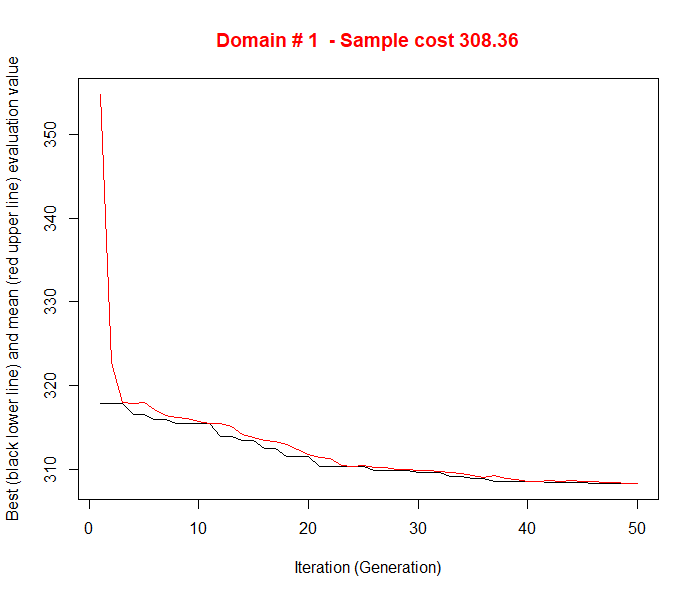}
		\caption{Optimization step for 4 metals concentration with 5 strata}
		\label{multi_optim}
	\end{figure}
	
	In listing~\ref{ssmulti} the structure of the optimized strata is reported. The total sample size is 219. The first target estimate (cadmium concentration) is the one that contributes most to the overall sample size: in fact its expected CV is just on the limit of 5\% set by the precision constraints. 
	
	\begin{lstlisting}[language=R]
	expected_CV(solution$aggr_strata)
	# cv(Y1) cv(Y2) cv(Y3) cv(Y4)
	# 0.05  0.028  0.033  0.033
	\end{lstlisting}
	
\begin{lstlisting}[basicstyle=\tiny,float=h,frame=tb,caption=\textit{SamplingStrata} optimized strata (multivariate case),label=ssmulti]
  Domain Stratum Population Allocation SamplingRate Lower_X1  Upper_X1 Lower_X2  Upper_X2
1      1       1        804         78     0.096891 0.000000  1.446959  0.00000  38.97286
2      1       2        965         92     0.095770 0.000000  3.802026 15.50852  73.20976
3      1       3        723         74     0.101837 1.175621  5.916244 28.05583  68.40705
4      1       4        437         46     0.105043 3.479107  8.731853 34.25202  80.43462
5      1       5        174         19     0.106353 5.440594 12.303164 46.90210 120.70553
   Lower_X3 Upper_X3 Lower_X4  Upper_X4
1   0.00000  89.1031   0.0000  177.3007
2  46.84002 170.2959 177.4111  380.9390
3 107.15412 249.7760 381.4216  649.5135
4 135.71557 317.1282 649.6953  906.9500
5 219.99896 462.6618 908.4930 1537.8878
\end{lstlisting}
	
\begin{figure}[H] 
		\centering
		\includegraphics[width=0.7\textwidth]{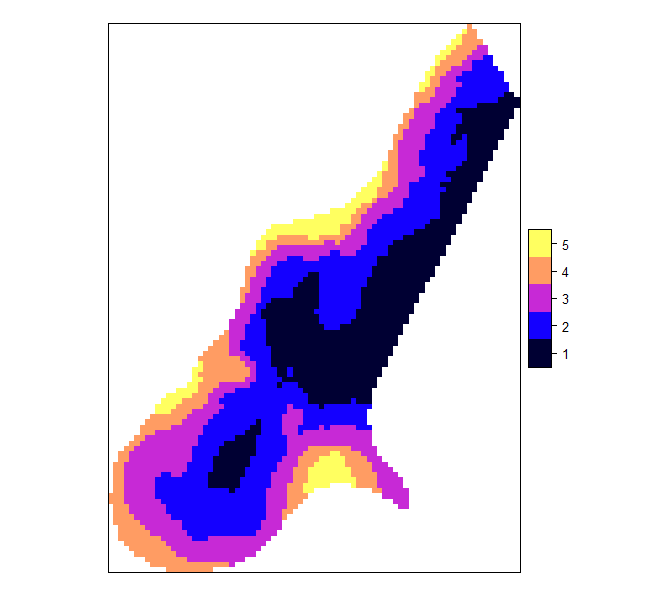}
		\caption{Final strata in multivariate case}
		\label{meusemulti}
	\end{figure}

	The resulting strata are visualized in Figure \ref{meusemulti}
	
	\subsection{Case study 2: Census Enumeration Areas Frame}
	\label{case2}
	
\begin{figure}[h!]
		\begin{subfigure}{.54\textwidth}
			\centering
			\includegraphics[width=1\linewidth]{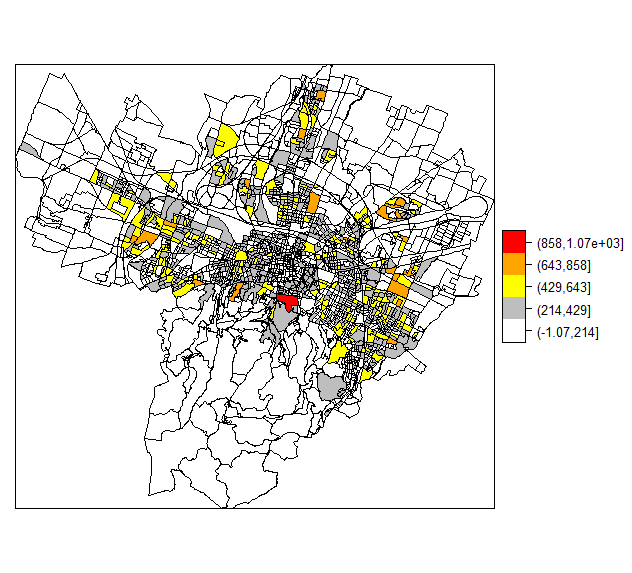}
			\caption{Population}
			\label{Bologna_P1:sfig1}
		\end{subfigure}%
		\begin{subfigure}{.5\textwidth}
			\centering
			\includegraphics[width=1\linewidth]{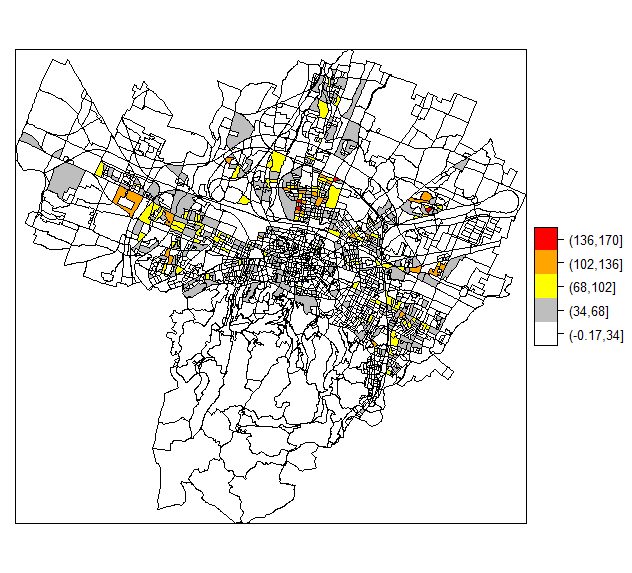}
			\caption{Foreign residents}
			\label{Bologna_P1:sfig2}
		\end{subfigure}
		\caption{Territorial distribution in the 2,234 Bologna EA's}
		\label{Bologna_P1}
	\end{figure}

	In this case study we will consider a sampling frame containing Census Enumeration Areas (EA's) as selection units: in particular, the 2,234 EA's included in Italian Municipality of Bologna. In this basic application, we consider as associated information only the total amount of population and of foreign residents. Population P is continuously updated by the Population Register, but here we use the last 2011 Population Census data. Territorial distribution is shown in Figure \ref{Bologna_P1}.
	
	The setting for the study is the following:
	
	\begin{itemize}
		\item the aim is to design a sample survey whose target is to produce estimates on the size of a given sub-population in the same Bologna territory, i.e. those of the foreign residents;
		\item data on sub-populations are available with a temporal lag (for instance, at the last Census) or on a subset of EA's (by means of a non optimized sample): in this case, we consider the second option, in line with the Istat Permanent Census, and will draw a simple random sample with sampling rate equal to 0.2;
		\item using data on total population (P1) and foreign residents (ST1) for the EA's where they are jointly available, it is possible to estimate different models, in particular:
		\begin{enumerate}
			\item Linear Model;
			\item Kriging;
			\item Spatial Linear Model;
		\end{enumerate}
		\item having set precision constraints on one or more target estimates of the sub-population of interest, the best stratification of the frame is obtained by running three different optimization steps, one for each of the above models;
		\item a compared evaluation of the three different solutions is obtained by simulating the selection of 1,000 samples, calculating for each of them the target estimates and finally deriving the coefficients of variation.
	\end{itemize}

	\subsubsection{Application on real data}
	
	\textit{Linear Model}
	\\
	
	\begin{lstlisting}[language=R]
	lm_1 <- lm(ST1 ~ P1,data=spoints_samp)
	summary(lm_1)
	\# Coefficients:
	\#   Estimate Std. Error t value            Pr(>|t|)    
	\# (Intercept) -0.053350   0.982107  -0.054               0.957    
	\# P1           0.118006   0.004238  27.844 <0.0000000000000002 ***
	\#   ---
	\# Residual standard error: 15.41 on 465 degrees of freedom
	\# Multiple R-squared:  0.6251,	Adjusted R-squared:  0.6243 
	
	\end{lstlisting}

	\begin{figure}[h!]
		\begin{subfigure}{.5\textwidth}
			\centering
			\includegraphics[width=1\linewidth]{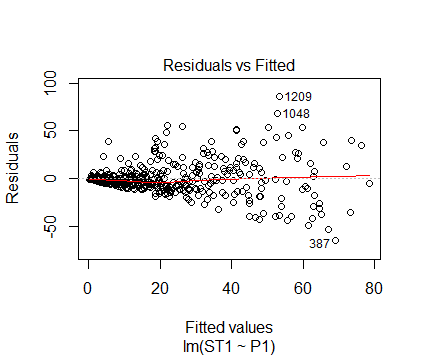}
			\caption{}
			\label{lm1:sfig1}
		\end{subfigure}%
		\begin{subfigure}{.5\textwidth}
			\centering
			\includegraphics[width=1\linewidth]{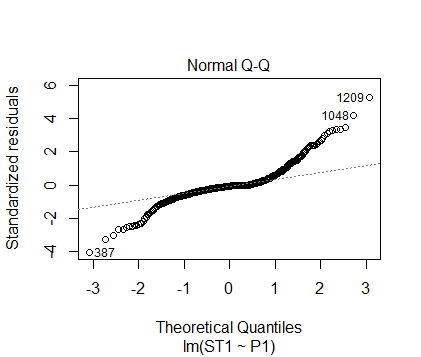}
			\caption{}
			\label{lm1:sfig2}
		\end{subfigure}
		\caption{Linear model between foreign residents sub-population and total population}
		\label{lm1_real}
	\end{figure}
	
	As for the distribution of residuals, see Figure \ref{lm1_real}. It is quite evident the presence of heteroscedasticity and the non-normality of residuals distribution. For this reason, we proceed to compute the heteroscedasticity index and related variance for each one of the three different models, and give corresponding values as input to the optimization steps. 
	
\begin{figure}[h!]
		\centering
		\includegraphics[width=0.8\linewidth]{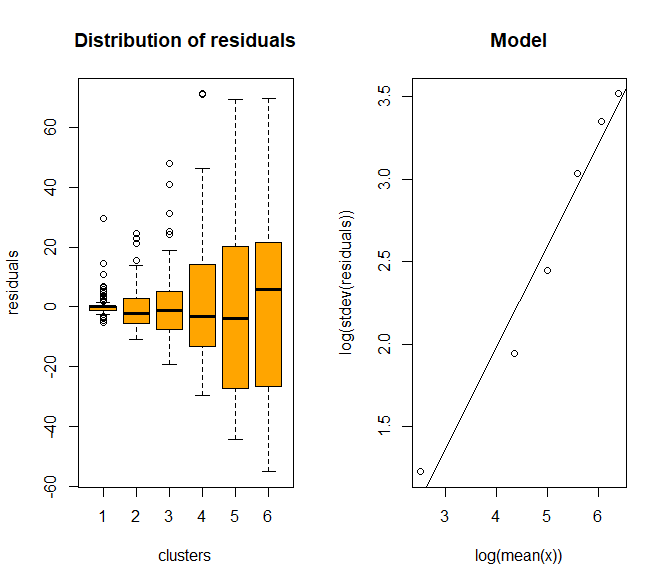}
		\caption{}
		\caption{Distribution of residuals and heteroscedasticity model}
		\label{gamma}
	\end{figure}

	We calculate the heteroscedasticity index and related variability accordingly to the Linear Model (see Figure \ref{gamma}):
	\begin{lstlisting}[language=R]
	gamma_sigma_1 <- computeGamma(e=summary(lm_1)$residuals,
	x=spoints_samp@data$P1,nbins=6)
	gamma_sigma_1
	# gamma     sigma  r.square 
	# 0.6146480 0.6210120 0.9662238 
	\end{lstlisting}
	
	where the first value is the heteroscedasticity index, and the second is the standard deviation 
	\\
	
	\newpage
	
	We build the sampling frame in this way: 
	
	\begin{lstlisting}[language=R]
	id    X1  Y1 domainvalue      lon      lat      ST1  P1
	1164  64  64           1 686797.2  4928305        8  64
	300 192 192           1 686739.8  4932537       42 192
	1791 190 190           1 684180.3  4929655       27 190
	309 350 350           1 687154.7  4932146      131 350
	895   0   0           1 682919.1  4931359        0   0
	1750 599 599           1 690758.9  4929418      107 599
	\end{lstlisting}
	
	that is, both stratification variable X1 and target variable Y1 have been set equal to the population total P1. 
	
	Having fixed a precision constraint of 3\% on the target variable, the first optimization is the one making direct use of the linear model, using the method \textit{continuous}:
	
	\begin{lstlisting}[language=R]
	model <- NULL
	model$type[1] <- "linear"
	model$beta[1] <- summary(lm_1)$coefficients[2]
	model$gamma[1] <- gamma_sigma_1[1]
	model$sig2[1] <- gamma_sigma_1[2]^2
	model <- as.data.frame(model)
	model
	# type      beta    gamma      sig2
	# 1 linear 0.1180064 0.614648 0.3856559
	
	set.seed(1234)
	solution1 <- optimStrata (
	method="continuous",
	errors=cv, 
	framesamp=frame,
	model=model,
	nStrata = 5)
	\end{lstlisting}
	
	This solution yields a sample size of 400. We now proceed to draw 1,000 samples and calculate the real CV for both the Y1 variable and for the real target variable, ST1:
	
	\begin{lstlisting}[language=R]
	framenew <- solution1$framenew
	outstrata <- solution$aggr_strata
	framenew$Y2 <- framenew$TARGET
	val1 <- evalSolution(framenew, outstrata, nsampl=1000)
	val1$coeff_var
	# CV1    CV2  dom
	# 1 0.0094 0.029 DOM1
	\end{lstlisting}
	
	The CV related to P1 is much lower that the required 3\%, but we see that the CV related to our true target variable (ST1) is compliant (2.9\%). 
	\\
	
	\textit{Kriging}
	\\
	
	We move now to the approach based on the use of Kriging.
	First, we obtain the model in this way:
	
	\begin{lstlisting}[language=R]
	# Variogram
	v <- variogram(ST1 ~ P1, data=spoints_samp, cutoff=3000, width=3000/30)
	# Estimation of psill, range and nugget with automap
	fit.vgm = autofitVariogram(ST1 ~ P1, spoints_samp, 
	model = c("Exp", "Sph", "Mat" ))
	plot(v, fit.vgm$var_model)
	fit.vgm$var_model
	# model    psill    range kappa
	# 1   Nug 114.3363    0.000   0.0
	# 2   Mat 135.2903 1232.796   0.2
	# Prediction 
	g <- gstat(NULL, "v", ST1 ~ P1, spoints_samp)
	v <- variogram(g)
	v.fit <- fit.lmc(v, g, 
	vgm(psill=fit.vgm$var_model$psill[2], 
	model=fit.vgm$var_model$model[2], 
	range=fit.vgm$var_model$range[2], 
	nugget=fit.vgm$var_model$psill[1]))
	preds <- predict(v.fit, Comune_BO_geo)
	\end{lstlisting}
	
	We calculate the heteroscedasticity index and standard deviation of errors in this way:
	
	\begin{lstlisting}[language=R]
	gamma_sigma_2 <- computeGamma(e=lm_pred$residuals[camp],
	x=frame1$P1[camp],
	nbins=6)
	gamma_sigma_2
	# gamma     sigma  r.square 
	# 0.5100489 0.9401730 0.9394493 
	\end{lstlisting}
	
	and calculate prediction errors in this way:
	
	\begin{lstlisting}[language=R]
	frame$var1 <- gamma_sigma_2[2]^2 * (frame1$P1^(2*gamma_sigma_2[1]))
	\end{lstlisting}	
	
	Once assigned to the frame the predicted values and the prediction errors to each unit in the frame, we calculate the fitting of the model in this way:
	
	\begin{lstlisting}[language=R]
	# Compute fitting
	lm_pred <- lm(ST1 ~ pred,data=frame)
	summary(lm_pred)$r.squared
	# [1] 0.6351977
	\end{lstlisting}
	
	We can now proceed with the optimization step using the method \textit{spatial}:
	
	\begin{lstlisting}[language=R]
	frame$var1 <- preds$v.var
	solution2 <- optimStrata (
	method="spatial",
	errors=cv, 
	framesamp=frame,
	nStrata = 5,
	fitting = summary(lm_pred)$r.squared, 
	range = fit.vgm$var_model$range[2],
	kappa = 1)
	\end{lstlisting}
	
	The sample size is now 330, much lower than the previous one, and the expected CV on the target variable is slightly above the desired 3\%, as it is 3.2\%. Once equalized (that is, once increased the sample size to the same one of the \textit{Linear Model}) it reduces to 2.85\%.
	\\
	
	\textit{Spatial Linear Model}
	\\
	
	Finally, we make use of third approach, the one based on the Spatial Linear Model:
	
	\begin{lstlisting}[language=R]
	lm_2 <- lm(ST1 ~ P1 + P1W, data=frame[camp,])
	summary(lm_2)
	# Coefficients:
	#   Estimate Std. Error t value            Pr(>|t|)    
	# (Intercept)  0.231107   1.457822   0.159               0.874    
	# P1           0.118487   0.004616  25.667 <0.0000000000000002 ***
	# P1W         -0.002018   0.007638  -0.264               0.792    
	# Residual standard error: 15.42 on 464 degrees of freedom
	# Multiple R-squared:  0.6251,	Adjusted R-squared:  0.6235 
	\end{lstlisting}
	
	We observe that the value of the coefficient related to P1W is very small, thus indicating a limited presence of spatial auto-correlation. 
	
	We calculate heteroscedasticity and variability:
	
	\begin{lstlisting}[language=R]
	gamma_sigma_3 <- computeGamma(e=(frame$target[camp] - 
	predict(lm_2,data=frame[camp,])),
	x=spoints_samp@data$P1,nbins=6)
	# gamma     sigma  r.square 
	# 0.6132473 0.6257764 0.9654370 
	\end{lstlisting}
	
	and calculate the value of the range:
	
	\begin{lstlisting}[language=R]
	v2 <- variogram(res_spatial  ~ 1, data=spoints_samp)
	fit.vgm2 = autofitVariogram(res_spatial  ~ 1, spoints_samp, 
	model = c("Exp","Sph","Mat"))
	fit.vgm2$var_model
	# model    psill    range kappa
	# 1   Nug 114.3825    0.000   0.0
	# 2   Mat 135.2430 1232.444   0.2
	\end{lstlisting}
	
	We can now proceed with the optimization step:
	
	\begin{lstlisting}[language=R]
	model
	# type      beta        beta2      sig2    range     gamma   fitting
	# 1 spatial 0.1184873 -0.002018293 0.3915961 1232.444 0.6132473 0.6251384
	
	solution3 <- optimStrata (
	model="continuous",
	errors=cv, 
	framesamp=frame,
	model=model,
	nStrata = 5)
	\end{lstlisting}
	
	obtaining a total allocation of 393. As in the case of the Linear Model, the expected CV on the target is compliant with the precision requirement. Once equalized, it reduces to 2.89\%.
	
\begin{figure}[H]
		\begin{subfigure}{.5\textwidth}
			\centering
			\includegraphics[width=0.7\linewidth]{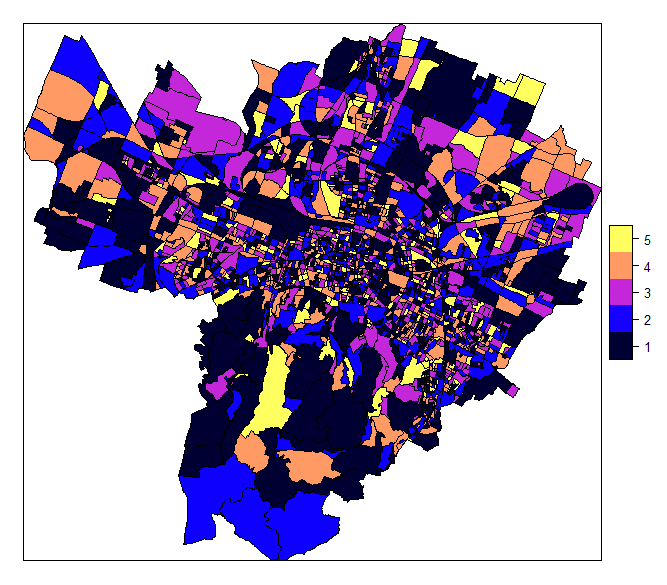}
			\caption{Stratified EA's}
			\label{bologna_real:sfig1}
		\end{subfigure}%
		\begin{subfigure}{.5\textwidth}
			\centering
			\includegraphics[width=0.7\linewidth]{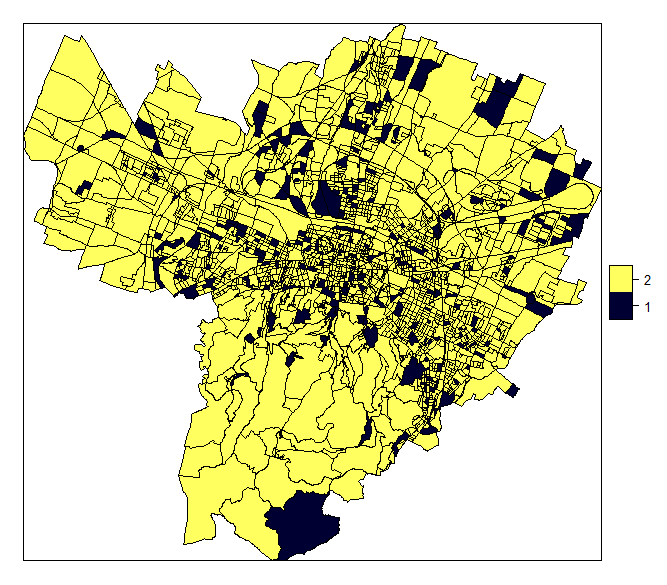}
			\caption{Sampled EA's}
			\label{bologna_real:sfig2}
		\end{subfigure}
		\caption{Stratification obtained with the Kriging Model}
		\label{bologna_real}
	\end{figure}

	As an example of a resulting stratification, we report the one obtained with the Kriging Model in Figure~\ref{bologna_real}, together with the 330 sampled EA's.
	
	The structure of the obtained strata, in terms of boundaries, together with allocation and sampling rate, is reported in Listing~\ref{reportRealPop}.
	
\begin{lstlisting}[float=h!,frame=tb,caption=Results of the application of the three models,label=reportRealPop]

 *** Linear model (gamma/sigma/R2 =  0.614648 0.621012 0.9662238 ) ***
R squared:  0.625082
Sample size 400
  Domain Stratum Population Allocation SamplingRate Lower_X1 Upper_X1
1      1       1        689         28     0.040516        0       36
2      1       2        617         82     0.133399       37      137
3      1       3        433         92     0.212173      138      248
4      1       4        421        126     0.300094      250      432
5      1       5        174         72     0.414277      433     1072
  CV(P1)  CV(ST1)
  0.0094 0.029 

 *** Kriging (gamma/sigma/R2 =  0.5006069 0.9888317 0.9372327 ) ***
R squared:  0.6351977
Sample size 330
  Domain Stratum Population Allocation SamplingRate Lower_X1 Upper_X1
1      1       1        685         37     0.054042     0.00     4.12
2      1       2        534         57     0.107322     4.15    13.94
3      1       3        467         75     0.161391    13.97    27.98
4      1       4        428         93     0.218032    28.04    49.32
5      1       5        220         68     0.307086    49.49   123.91
  CV(P1)  CV(ST1)
  0.0105 0.0321 
... with the same sample size than Linear Model
  CV(P1)  CV(ST1)
  0.0088 0.0285 

 *** Spatial model (gamma/sigma/R2 =  0.6241951 0.5883335 0.9638095 ) ***
R squared:  0.6251384
Sample size 393
  Domain Stratum Population Allocation SamplingRate Lower_X1 Upper_X1
1      1       1        689         29     0.042281        0       36
2      1       2        617         81     0.131964       37      137
3      1       3        433         88     0.203478      138      248
4      1       4        421        123     0.291982      250      432
5      1       5        174         72     0.413850      433     1072
  CV(P1)  CV(ST1)
  0.009 0.0304 
... with the same sample size than Linear Model
  CV(P1)  CV(ST1)
  0.0095 0.0289 
\end{lstlisting}

	\subsubsection{Simulation}
	
	The results obtained in the application to real data indicate that:
	
	\begin{itemize}
		\item both \textit{Linear Model} and \textit{Spatial Linear Model} are compliant with the precision requirement of 3\%, while \textit{Kriging} is slightly over;
		\item conversely, in terms of sample size, the \textit{Kriging} is the one with the minimum;
		\item once equalized (that is, after rendering equal all the sample sizes of the three solutions), the expected CVs are about the same, with a minimum for the \textit{Kriging};
		\item the structure of the obtained optimized strata varies from one solution to another, both in terms of strata boundaries and allocations.
	\end{itemize}
	
	The substantial equivalence of the results obtained with three models is probably due to a weak spatial correlation of the variable of interest. 
	\\
	
	In order to analyse the contribution of spatial component in the overall variance to results obtainable by different models, we decided to carry out a simulation by generating for each EA in the frame an artificial variable $Z$, characterized by a high spatial correlation, by making use of the following equation: 
	\\
	
	$Z = Y \beta_{1} + W Y \beta_{2} + u\cdot Y^{\gamma}$
	\\
	
	In our case $Y$ is the population P1 in each EA, $\beta$ has been set equal to 1, $W$ is the weighted contiguity matrix, and the errors $\epsilon$ are spatially correlated and affected by heteroscedasticity.  
	The error term $u$ has been generated as a \textit{Gaussian random field}, whose parameters are the following:
	\begin{enumerate}
		\item the covariance model is exponential;
		\item the error variance has been set equal to 2000;
		\item the range of spatial correlation has been set to 2000;
		\item the values of heteroscedasticity index ($\gamma$) ranges from 0 to 0.7 with a step equal to 0.05.
	\end{enumerate}
	
\begin{figure}[H]
		\begin{subfigure}{.5\textwidth}
			\centering
			\includegraphics[width=1.2\linewidth]{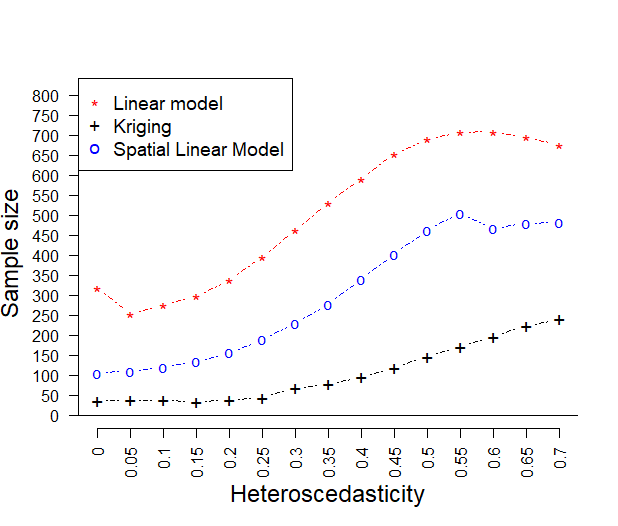}
			\caption{Performance in terms of sampling size}
			\label{simulations:sfig1}
		\end{subfigure}%
		\begin{subfigure}{.5\textwidth}
			\centering
			\includegraphics[width=1.2\linewidth]{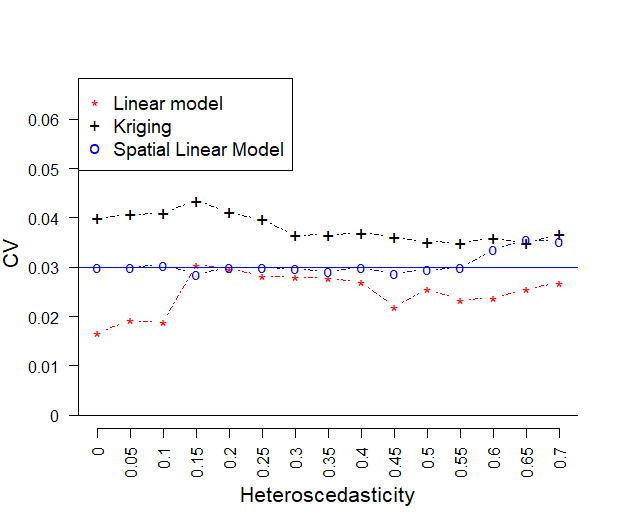}
			\caption{Performance in terms of CV on target variable}
			\label{simulations:sfig2}
		\end{subfigure}
		\begin{subfigure}{.5\textwidth}
			\centering
			\includegraphics[width=1.5\linewidth]{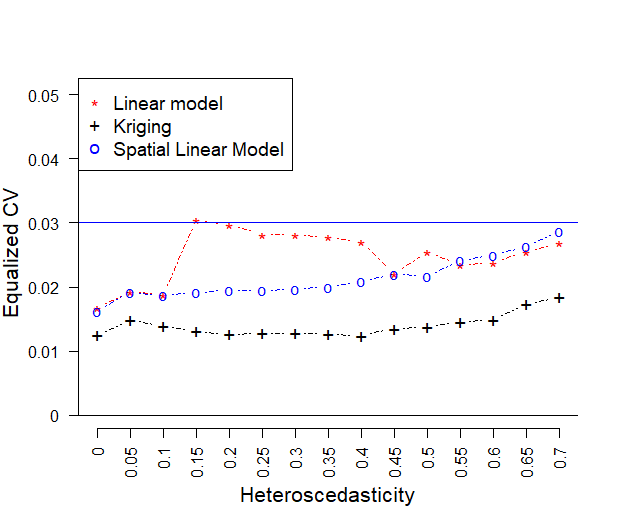}
			\caption{Performance in terms of equalized CV on target variable}
			\label{simulations:sfig3}
		\end{subfigure}
		\caption{Results of simulation}
		\label{simulations}
	\end{figure}

	In Table \ref{sim_res} and in Figure~\ref{simulations} the results of the simulation are reported.
	
	\begin{table}[H]
		\centering
		\begin{tabular}{||c|c|c|c|c|c|c|c|c|c||}
			\hline 
			\cline{3-10}
			& & \multicolumn{2}{|c|}{Linear} & \multicolumn{3}{|c|}{Kriging} & \multicolumn{3}{|c||}{Spatial}
			\\
			Run & Heter.lev.& n  & CV & n & CV & CV eq.  & n & CV & CV eq.  \\ 
			\hline 
			1  &   0  &319  &0.0167   & 36	&0.0399 &0.0125  &105  & 0.0299 &0.0162\\ 
			2  &0.05  &254  &0.0192   & 37	&0.0407 &0.0148  &110  & 0.0300 &0.0192\\ 
			3  &0.10  &275  &0.0188   & 37	&0.041; &0.0139  &120  & 0.0304 &0.0187\\ 
			4  &0.15  &297  &0.0305   & 34	&0.0435 &0.0131  &134  & 0.0286 &0.0186\\ 
			5  &0.20  &338  &0.0297   & 38	&0.0411 &0.0126  &156  & 0.0300 &0.0195\\ 
			6  &0.25  &395  &0.0282   & 44	&0.0397 &0.0128  &188  & 0.0300 &0.0194\\ 
			7  &0.30  &462  &0.0281   & 67	&0.0365 &0.0128  &230  & 0.0301 &0.0194\\ 
			8  &0.35  &530  &0.0278   & 78	&0.0366 &0.0127  &276  & 0.0292 &0.0200\\ 
			9  &0.40  &591  &0.0270   & 95	&0.037; &0.0124  &341  & 0.0297 &0.0206\\ 
			10 &0.45  &653  &0.0220   &119	&0.0361 &0.0135  &402  & 0.0287 &0.0220\\ 
			11 &0.50  &690  &0.0255   &146  &0.0352 &0.0137  &456  & 0.0286 &0.0220\\ 
			12 &0.55  &707  &0.0234   &171  &0.0349 &0.0146  &505  & 0.0299 &0.0241\\ 
			13 &0.60  &709  &0.0238   &196  &0.036; &0.0149  &467  & 0.0336 &0.0250\\ 
			14 &0.65  &695  &0.0255   &223  &0.0349 &0.0173  &478  & 0.0344 &0.0279\\ 
			15 &0.70  &676  &0.0268   &240  &0.0368 &0.0184  &483  & 0.0353 &0.0287\\ 
			\hline 
		\end{tabular} 
		\caption{Simulation results.}
		\label{sim_res}
	\end{table}

	We see that:
	
	\begin{enumerate}
		\item as for the overall sample sizes required by the three models (see Figure \ref{simulations:sfig1}), those obtained by the \textit{Kriging} are always the minimum, followed by the \textit{Spatial Linear Model}, while the \textit{Linear Model} is characterized by very high levels of smple size;
		\item in terms of resulting expected coefficient of variation on the target variable (Figure \ref{simulations:sfig2}), the \textit{Spatial Linear Model} is the only one whose expected CV are in line with the precision requirement; the \textit{Linear Model} is the one that yields the minimum values, while the \textit{Kriging} produces the highest values; 
		\item finally, considering the equalized expected CV (the CV that is obtained when the sample size of \textit{Kriging} and \textit{Spatial Linear Model} are increased to the one of the \textit{Linear Model}), the \textit{Kriging} is always the best, while the other two are characterized by near the same values. 
	\end{enumerate}
	
	As a first indication, we can conclude that:
	\begin{itemize}
		\item if the aim is to minimize the overall sample size, \textit{Kriging} is the right choice: there is no warranty that the precision constraint will be respected, but the efficiency of the solution is maximized;
		\item if conversely we want to stick to the precision constraint, the \textit{Spatial Linear Model} should be preferred, because it has to be expected that the sample size is the one actually needed in order to be compliant.
	\end{itemize}	
	
	\section{Conclusions and future work}
	\label{conclusions}
	
	The functionalities of the package \textit{SamplingStrata} have been enriched by new developments, having the aim to increase its effectiveness and efficiency, and to extend its applicability. 
	
	The changes in the Genetic Algorithm when method "\textit{atomic}" is used, respond to the need of increasing its computational performance. 
	
	The new method "\textit{continuous}" allows to produce strata that are more interpretable and manageable, and in some cases also more efficient. 
	
	The handling of \textit{anticipated variance} permits to compute variance in strata in a more correct way, taking into account the difference between available variables in the frame, and the variables that are the real target of the sampling survey. Very important to this aim is the consideration of heteroscedasticity of residuals, and the new function \textit{computeGamma} allows to estimate its value.
	
	The extension to the case where units in the frame can be georeferenced or geocoded covers the more and more important field of spatial sampling. The two alternatives provided, i.e. (i) the extension of model types to the 'spatial' linear model in the handling of anticipated variance, and (ii) the development of a method, "\textit{spatial}", both allow to take into account the spatial auto-correlation when calculating variance in strata, thus increasing the correctness and efficiency of optimized solutions. In the reported case studies, the applicability of these new functionalities has been illustrated, though there is not yet a clear evidence of what may be the best one. In our opinion, we need to investigate better the use of the different variants of the \textit{Spatial Linear Model}. 
	
	Another required development will concern sample selection techniques peculiar of spatial sampling. Once optimal strata have been defined, it would be important to select units inside them by using techniques (as, for instance, grid sampling), that ensure to harness the spatial auto-correlation at the maximum extent.

	\section*{Acknowledgement}
	
	We would like to thank Mervyn O’Luing, Steven Prestwich and S. Armagan Tarim for their proposals for improving the efficiency of the Genetic Algorithm at the basis of the optimization step; and Jaap De Gruijter for his clarifications regarding variance calculation including spatial co-variance.
	
	\pagebreak
	
	\baselineskip 10pt
	\bibliography{SS}
	\bibliographystyle{model5-names}
	
\end{document}